\documentstyle[aps,prl,epsf,multicol]{revtex}
\begin{document}
\title{The Local and the Occupation Time of a Particle Diffusing in
       a Random Medium}

\author{ Satya N. Majumdar$^1$ and Alain Comtet$^2,^3$}

\address{
{\small $^1$Laboratoire de Physique Quantique (UMR C5626 du CNRS), Universit\'e Paul
        Sabatier, 31062 Toulouse Cedex. France}\\
{\small $^2$Laboratoire de Physique Th\'eorique et Mod\`eles Statistiques,
        Universit\'e Paris-Sud. B\^at. 100. 91405 Orsay Cedex. France}\\
{\small $^3$Institut Henri Poincar\'e, 11 rue Pierre et Marie Curie, 75005 Paris, France}}
%\date{\today}
\maketitle

\begin{abstract}
We consider a particle moving in a one dimensional potential which has a symmetric deterministic part and a
quenched random part. We study analytically the probability distributions of the local time (spent by the
particle
around its mean value) and the occupation time (spent above its mean value) within an observation
time window of size $t$. In the absence of quenched randomness, these
distributions have three typical asymptotic behaviors depending on whether the deterministic
potential is unstable, stable or flat. These asymptotic behaviors are shown to get drastically modified
when the random part of the potential is switched on leading to the loss of self-averaging
and wide sample to sample fluctuations.
\end{abstract}


\begin{multicols}{2}
How many days, out of a total number of $t$ days, does a tourist spend in a given place?
If the tourist is a Brownian particle, this local time (per unit volume) $n_t(\vec r)$ spent by it in an
infinitesimal neighbourhood of a point $\vec r$ in space has been of interest to physicists
and mathematicians for decades.
If $\vec R (t')$ denotes the position of the random walker at time
$t'$, the local time is defined as
\begin{equation}
n_t(\vec r)=\int_0^t \delta\left[
{\vec R} (t')-\vec r\right]dt'.
\label{deflt}
\end{equation}
Clearly $\int n_t(\vec r)d\vec r=t$. The local time is a very useful quantity with a variety
of important applications in fields ranging from physics to biology. For example, in the context of
polymers in a solution, the local time $n_t(\vec r)$ is proportional to the concentration
of monomers at $\vec r$ in a polymer of length $t$ and can be measured via light
scattering experiments. In ``True" self-avoiding walk (TSAW), a simple model of a polymer that takes into
account the excluded volume effect, the local time already spent by the walker at a site decides the future
rate of revisits to that site\cite{APP}. The
distribution of local time (LTD) also plays an important role in the study
of diffusion in porous rocks\cite{MdM,BG}. Another important application of the local time is in bacterial
chemotaxis where the phenomenon of `tumbling'
is quantified by the local time spent by a bacteria at a point\cite{Koshland}.

A related quantity, also of wide interest to both physicists\cite{GL} and
mathematicians\cite{math}, is the
occupation time $T_t(D)=\int_D n_t(\vec r)d\vec r$ spent by the
walker in a given region $D$ of space.
Recently the study of the occupation time has seen a revival due to its newfound applications
in the context of persistence in nonequilibrium statistical physics\cite{Review}.
The dynamics in such systems is typically modelled by a stochastic process $x(t)$
whose statistical properties provide important informations about the history of evolution
in these systems\cite{Review}. A quantity that acts as a useful probe to this
history dependence is the occupation time $T_t=\int_0^t
\theta\left[x(t')\right]dt'$, the time spent by the process on the
positive side within a window of size $t$\cite{Occup}.
For instance, if $x(t)$ represents a one dimensional
Brownian motion, the probability distribution of the occupation time (OTD) is
given by the celebrated
`arcsine' law of L\'evy\cite{Levy},
${\rm Prob}[T_t=T]=1/{\pi \sqrt{T(t-T)}}$ which diverges at the end points
$T=0$ and $T=t$
indicating the `stiffness' of the Brownian motion, i.e., a Brownian path starting at a positive
(negative) point
tends to remain positive (negative). In the mathematics literature this
result has been extended in several directions using modern tools of
probability theory such as martingale calculus and excursion theory\cite{Y}.
The OTD has been studied in a variety of physical systems. For example, it has
been proposed as a useful tool to analyse the morphological dynamics of interfaces\cite{TND}.
Recently the OTD has also been used to analyse the experimental data on
the ``on-off" fluorescence intermittency
emitting from colloidal (CdSe) semiconductor quantum dots\cite{ENS}.
Exotic properties such as a phase transition
in the ergodicity of a process by tuning a parameter
have also been exhibited by the OTD in the diffusion equation\cite{NL}.

While the LTD and the OTD have been studied extensively for pure systems, they have
received much less attention in systems with quenched disorder, the subject of this Letter. The
study of these quantities can provide valuable informations in disordered systems. For example, if one
launches a tracer particle in a system with localized impurities, the particle diffuses in this
random medium, occasionally gets pinned in the region near the impurities till the thermal
fluctuations lift it out of the local potential well and then it diffuses again. Since the local
time spent by the particle at a given point in space is related to the concentration of impurities
there, it can be used as a valuable probe to image the inhomogeneities in a given sample.
In this Letter we present exact results for the LTD and the OTD
in the Sinai type models\cite{Sinai} where
a particle diffuses in a one dimensional random potential. Both of these distributions turn out to
have very rich behaviors despite the simplicity of the model. The problem of a particle diffusing in a
random potential, apart from being a basic problem in
disordered systems, has
numerous applications in
various physical processes\cite{BG,BCGL} including the diffusion of electrons in disordered medium,
glassy activated
dynamics of dislocations in solids, dynamics of random field magnets, dynamics near the helix-coil
transitions in heteropolymers and more recently the dynamics of denaturation of a single DNA molecule
under external force\cite{LN}. Hence one expects that
the knowledge of the LTD and the OTD in the simple Sinai type models, derived here, will have
broader applications in several of these quenched disordered systems.

We start with the Langevin equation of motion of an overdamped particle
\begin{equation}
{dx \over {dt}}= F(x) + \eta(t),
\label{lange1}
\end{equation}
where $\eta(t)$ is a thermal Gaussian white noise with zero mean and a correlator
$\langle \eta(t)\eta(t')\rangle =2k_BT \delta(t-t')$. For simplicity we set $k_BT=1$. The force
$F(x)=-dU/dx$ is
derived from a potential which has a deterministic and a random part,
$U(x)=U_d(x) +U_r(x)$. In the continuous version of the Sinai model we choose the random potential $U_r(x)
=\sqrt{\sigma}\int_0^x \xi(x')dx'$ to be a Brownian motion in space where $\xi(x)$ is a quenched
Gaussian noise with zero mean and a correlator $\langle \xi(x)\xi(x')\rangle =\delta(x-x')$.
The goal is to first compute the probability distribution of the local time,
$n_t(a)=\int_0^{t}\delta\left(x(t')-a\right)dt'$ and that of the occupation time
$T_t(a)=\int_0^{t}\theta\left(x(t')-a\right)dt'$ corresponding to level $a$ for a given sample of
quenched disorder and then obtain the disorder averaged distributions.
For simplicity, we will consider $U_d(x)$ to be symmetric so that
the mean position of
the particle is at zero and restrict ourselves to study the distributions of $n_t=n_t(0)$
and $T_t=T_t(0)$ corresponding to the natural choice of the level $a=0$. However
our results are easily generalizable to more general potentials and to arbitrary levels
$a$.

It turns out that the generic asymptotic scaling behaviors of the LTD and the OTD, at a
qualitative level, depend on whether the deterministic potential $U_d(x)$ is
unstable ($U_d(x)\to
-\infty$
as $x\to \pm \infty$), stable ($U_d(x)\to \infty$ as $x\to \pm \infty$) or flat ($U_d(x)=0$). Quantitatively
however, the LTD and the OTD do depend on the details of the potential $U_d(x)$. To keep
the discussion simple, we present explicit results here for the case when $U_d(x)=-\mu |x|$,
even though our techniques can be extended to other potentials as well. Thus we will consider
Eq. (\ref{lange1}) with the force $F(x)= \mu\, {\rm sign}(x) + \sqrt{\sigma}\,\xi(x)$.
For $\mu>0$, the deterministic force is repulsive from the origin and for $\mu<0$, the force is
attractive.
For $\mu=0$, Eq. (\ref{lange1}) reduces exactly to the Sinai model.

It is useful to start with a generalized variable $\tau_t=\int_0^t V\left[x(t')\right]dt'$ with
an arbitrary functional $V\left[x(t')\right]$ which reduces to the local time $n_t$
and the occupation time $T_t$ when $V(x)=\delta(x)$ and $V(x)=\theta(x)$ respectively.
Let $P_x(T,t)$ be the probability that $\tau_t=T$ given that the particle starts at $x(0)=x$.
The Laplace transform $Q_p(x,t)=\int_0^{\infty}e^{-pT}P_x(T,t)dT$ plays a crucial role in our
subsequent analysis. By definition $Q_p(x,t)={\langle e^{-p\int_0^t V\left[x(t')\right]dt'}\rangle}_x$
where ${\langle \rangle}_x$ denotes the average over all histories of the particle upto time $t$
starting at $x$ at $t=0$. Using the evolution equation (\ref{lange1}) it is straightforward to see
that $Q_p(x,t)$ satisfies, for arbitrary $F(x)$,  the backward Fokker-Planck equation
\begin{equation}
{{\partial Q_p}\over {\partial t}}={1\over {2}} {{\partial^2 Q_p}\over {\partial x^2}}+F(x){{\partial
Q_p}\over {\partial x}}-pV(x)Q_p,
\label{qpxt}
\end{equation}
with the initial condition $Q_p(x,0)=1$. After a further Laplace transform, now with respect to $t$,
$u(x)=\int_0^{\infty} e^{-\alpha t} Q_p(x,t)dt$ satisfies the equation
\begin{equation}
{1\over {2}} u^{''} + F(x)\,u^{'} -\left[\alpha+pV(x) \right]u=-1,
\label{upxa}
\end{equation}
where $u^{'}(x)=du/dx$ and we have suppressed the $\alpha$ and $p$ dependence of $u(x)$ for notational
convenience. So far the discussion is quite general. We now consider the local
and the occupation time separately.

{\it Local Time}\, : In this case $V(x)=\delta(x)$. We need to solve Eq. (\ref{upxa})
separately for $x>0$ and $x<0$ and then match the solutions at $x=0$. We write
$u_{\pm}(x)=1/\alpha + A_{\pm} y_{\pm}(x)$ where $y_{\pm}(x)$ satisfy the homogeneous equations
\begin{equation}
{1\over 2}y_{\pm}^{''}+F(x) y_{\pm}^{'}-\alpha y_{\pm}=0,
\label{yx}
\end{equation}
respectively in the regions $x>0$ and $x<0$ with the boundary conditions $y_{+}(x\to \infty)=0$
and $y_{-}(x\to -\infty)=0$. The constants
$A_{\pm}$ are determined from the matching
conditions, $u_{+}(0)=u_{-}(0)=u(0)$ and $u_{+}^{'}(0)-u_{-}^{'}(0)=2p u(0)$. Eliminating
the constants, we get $u(0)=\lambda(\alpha)/{\alpha [p+\lambda(\alpha)]}$ where
$\lambda(\alpha) =[z_{-}(0)-z_{+}(0)]/2$ and $z_{\pm}(x)=y_{\pm}^{'}(x)/y_{\pm}(x)$.
For simplicity, we will restrict ourselves only to $P_0(n_t=T,t)$ corresponding to the
natural choice of the starting point $x(0)=0$, though our methods can be easily generalized
to arbitrary initial positions.
Since $\lambda(\alpha)$ is independent of $p$, one can easily invert the Laplace transform $u(0)$ with
respect to $p$ to get
\begin{equation}
G(\alpha)=\int_0^{\infty}e^{-\alpha t}P_0(n_t=T,t)dt ={\lambda(\alpha)\over {\alpha}}e^{-\lambda(\alpha)T},
\label{ga}
\end{equation}
a general result valid for arbitrary $F(x)$.
To proceed more we choose $F(x)=\mu\, {\rm sign}(x)+\sqrt{\sigma}\xi(x)$ and consider
the implications of Eq. (\ref{ga}) for
the pure case $\sigma=0$
first for arbitrary $\mu$.

Solving Eqs. (\ref{yx}) for $F(x)=\mu\, {\rm sign}(x)$ we find $y_{\pm}(x)=y_{\pm}(0)e^{\mp \beta x}$
with $\beta=\mu+\sqrt{\mu^2+2\alpha}$. This gives $\lambda(\alpha)=\beta=\mu+\sqrt{\mu^2+2\alpha}$.
By analysing this Laplace transform one finds that
there are three different asymptotic (large $t$) behaviors of $P_0(T,t)$ depending on
whether the potential is unstable ($\mu>0$), stable ($\mu<0$) or flat ($\mu=0$). These asymptotic behaviors
for the particular potential chosen here ($U_d(x)=\mu|x|$) can be shown to be typical for generic potentials.

$\bullet$ {\it Unstable potential}\,($ \mu>0$)\, : In this case, the LTD approaches a steady state in the
large window size $t\to \infty$ limit, $P_0(T)=2\mu e^{-2\mu T}$ obtained by
taking $\alpha\to 0$ limit
in Eq. (\ref{ga}). Physically it indicates that for repulsive force the
particle eventually goes to either $\infty$ or $-\infty$ and occasionally
hits the origin according to a Poisson process. This asymptotic exponential distribution, $P_0(T)=\lambda(0)e^{-\lambda(0)T}$
is indeed universal (up to a rescaling factor of
time) for any unstable potential.

$\bullet$ {\it Stable potential}\, ($\mu<0$)\, : For generic stable potentials the system approaches a
stationary state
in the large $t$ limit and the stationary probability distribution $p(x)$ for the position of the particle
is given by the Gibbs measure, $p(x)=e^{-2U_d(x)}/Z$ where $Z=\int_{-\infty}^{\infty}e^{-2U_d(x)}dx$ is the
partition function.
Hence as $t\to \infty$, simple ergodicity arguments indicate that the local time $n_t \to
\int_0^t\langle
\delta\left[x(t')\right]\rangle dt'
\to p(0)\,t$, i.e., the LTD is simply $P_0(n_t=T,t)=\delta\left(T-p(0)t\right)$, a result that can also be
proved rigorously\cite{longpaper}. For example, for the example chosen here, one finds $p(0)=|\mu|$
and hence $P_0(T,t)=\delta(T-|\mu|t)$ as $t\to \infty$. It turns out that the approach to
this asymptotic delta function is rather slow in the following sense. Indeed
one finds that in the scaling limit
$t\to \infty$, $n_t\to \infty$ but keeping the ratio $n_t/t=y$ fixed,
the LTD behaves as, $P_0(n_t=T,t)\sim \exp\left[-t\Phi(T/t)\right]$ where the large deviation function (LDF)
$\Phi(y)$ is given from Eq. (\ref{ga}) via a steepest descent calculation
\begin{equation}
\Phi(y)= {\rm max}_{\alpha}\left[-\alpha+y\lambda(\alpha)\right],
\label{ld}
\end{equation}
which, once again, is a general result valid for any arbitrary stable potential. In our example,
substituting $\lambda(\alpha)=-|\mu|+\sqrt{\mu^2+2\alpha}$ in Eq. (\ref{ld}) and maximizing we get
the exact LDF, $\Phi(y)=(y-|\mu|)^2/2$. Thus in this particular example
$P_0(T,t)$ is a shifted Gaussian with mean at $T/t=|\mu|$ whose width decreases as a power law
$t^{-1/2}$ and only in the strict $t\to \infty$ limit, it approaches a delta function
peaked at $T/t=|\mu|$. This result also follows
if one assumes the validity of the
central limit theorem (CLT). However, for arbitrary stable potential, the CLT breaks down
and the LDF $\Phi(y)$ shows departure from the simple quadratic form\cite{longpaper,MB}.

$\bullet$ {\it Flat potential} ($\mu=0$)\, : In this case, one finds by inverting the Laplace tranform
in Eq. (\ref{ga}) that the LTD is Gaussian for all $T$ and $t$, $P_0(T,t)=\sqrt{2/{\pi t}}\,e^{-T^2/2t}$.

These behaviors for the pure system ($\sigma=0$) get drastically modified when the random potential is
switched on ($\sigma>0$). The equation (\ref{ga}) still remains valid for
each realization of $F(x)$. For instance by taking the $\alpha\to 0$
limit one may recover the expression of the thermal average of the local
time which was interpreted in\cite{BCGL} as a trapping time. Here our aim is
quite different since we want to
compute the disorder averaged LTD ${\overline {P_0(n_t=T,t)}}$. From Eq. (\ref{ga}), one needs to know the
distribution of $\lambda(\alpha)=[z_{-}(0)-z_{+}(0)]/2$ which is now a random variable since
$F(x)$ is random. It turns out the distribution of $\lambda(\alpha)$ can be computed exactly by adopting
techniques that have appeared before in the Sinai model in other contexts\cite{BCGL,CM,CD}. We defer
the technical details for a future publication\cite{longpaper} and present only the final results here.
One gets ${\overline {\exp[-\lambda(\alpha)T]}}=q^2(T)$ with
\begin{equation}
q(T)=(1+\sigma T)^{-\mu/{2\sigma}}  { {K_{\mu/\sigma}\left( {{\sqrt{2\alpha(1+\sigma T)}}\over
{\sigma}}\right)} \over
{ {K_{\mu/\sigma}\left( { {\sqrt{2\alpha}} \over
{\sigma} }\right) } } },
\label{qT}
\end{equation}
where $K_\nu(x)$ is the modified Bessel function of order $\nu$\cite{GR}.
Averaging Eq. (\ref{ga}) over disorder we finally get the exact formula
\begin{equation}
\int_0^{\infty}{\overline {P_0(T,t)}}e^{-\alpha t}dt=-{1\over {\alpha}}{d\over {dT}}[q^2(T)],
\label{dav}
\end{equation}
where $q(T)$ is given by Eq. (\ref{qT}). The asymptotic behaviors can be deduced by analysing
Eq. (\ref{dav}).

$\bullet$ $\mu>0$\, : In this case, we find that as
$t\to \infty$, ${\overline {P_0(T,t)}}$ tends to a steady
state distribution, ${\overline {P_0(T)}}=2\mu (1+\sigma T)^{-2\mu/\sigma-1}$
for all $T\ge 0$. Thus the
disorder averaged LTD
has a broad power law distribution even though for each sample the LTD has a narrow exponential
distribution. This indicates wide sample to sample fluctuations and lack of self averaging.

$\bullet$ $\mu<0$\, : In this case, for each sample of the disorder, the LTD tends to the delta
function $P_0(T,t)\to \delta\left(T-p(0)t\right)$ as discussed before. However, the Gibbs measure
$p(0)$ varies from sample to sample. On averaging over disorder (or equivalently the peak positions),
one finds a broad distribution for ${\overline {P_0(T,t)}}$. In the scaling
limit
$t\to \infty$, $n_t=T\to \infty$ but keeping the ratio $T/t$
fixed,
we find that ${\overline {P_0(T,t)}}\to {1\over {t}}f(T/t)$ where the scaling function can be computed
exactly by analysing Eq. (\ref{dav}),
\begin{equation}
f(y)= \left[ { {\sqrt{\pi}}\over
{\sigma(2\sigma^3)^{(\nu-1)/2}\Gamma^2(\nu)} }\right]y^{3(\nu-1)/2}e^{-y/\sigma}W_{\nu,\nu}(2y/\sigma),
\label{Whit}
\end{equation}
with $\nu=|\mu|/\sigma$ and $W_{\nu,\nu}(x)$ is the Whittaker function\cite{GR}. The scaling function
increases as
$f(y) \sim y^{\nu-1}$ for small $y$ and eventually decays for large $y$ as
$f(y)\sim y^{(4\nu-3)/2}e^{-2y/\sigma}$. Once again the disorder modifies the behavior
of the LTD rather drastically.

$\bullet$ $\mu=0$ (Sinai model)\,: In this case we find that for large $t$, ${\overline {P_0(n_t=T,t)}}\to
{1\over {t\log^2 t}}f_S(T/t)$ where $f_S(y)=2e^{-y/\sigma}K_0(y/\sigma)/y$. However this scaling breaks
down for very small $y$ when $y<<\sigma$.

We now turn to the OTD, $R_0(T,t)={\rm Prob}(T_t=T,t)$ given that the particle starts at $x=0$ at $t=0$. In
this case,
the double Laplace transform
$u_p(x)=\int_0^{\infty}dt e^{-\alpha t}\int_0^t e^{-pT}R_x(T,t)dT$ satisfies
Eq.(\ref{upxa})
with $V(x)=\theta(x)$. Since the rest of the calculations are very similar to the LTD case, we just
present the final results omitting the details.

$\bullet$ {\it Unstable potential}: Consider the pure case ($\sigma=0$) first. Since the
deterministic potential is symmetric, one has $R_0(T,t)=R_0(t-T,t)$, i.e. the OTD is symmetric around
its mean value $t/2$. In the limit of a large window size $t\to \infty$, it turns out that the
part of the OTD to the left of the midpoint $T=t/2$ approaches a steady ($t$
independent) distribution $R_L(T)$ with the normalization $\int_0^{\infty}R_L(T)dT=1/2$. The right
half of the OTD, which carries an equal total weight $1/2$ is pushed to $\infty$ since the midpoint $t/2$
itself goes to $\infty$. This conclusion is valid for any symmetric deterministic potential. For the case
$F(x)=\mu\, {\rm {sign}}(x)$, we get explicitly, $R_L(T)=\mu^2e^{-u^2}\left[1-3\sqrt{\pi}ue^{9u^2}{\rm
{erfc}}(3u)\right]/{\sqrt {\pi}u}$ where $u=\mu{\sqrt {T/2}}$ and
${\rm {erfc}}(x)$ is the complementary
error function. Thus $R_L(T)\approx \mu\sqrt{2/{\pi T}}$ for small $T$ and decays exponentially
for large $T$, $R_L(T)\sim T^{-3/2}\,e^{-\mu^2 T/2}$. When the disorder is switched on ($\sigma>0$),
this asymptotic behavior for
the pure case does not change qualitatively. Once again the left part of the disordered averaged OTD
tends to a $t$ independent form ${\overline {R_L(T)}}$. In fact, the small $T$
behavior of ${\overline
{R_L(T)}}\approx \mu\sqrt{2/{\pi T}}$ remains the same as in the pure case. However, for large
$T$, while the OTD still decays exponentially
${\overline {R_L(T)}}\sim e^{-bT}$, the decay coefficient $b$ turns out to be different
from the pure case and is given by the negative of the real part of the zero (closest to the origin) of
$K_{\mu/\sigma}(\sqrt{2z}/\sigma)=0$ in the complex $z$ plane.

$\bullet$ {\it Stable potential}: As in the case of the LTD we find that for the pure
case, for generic stable potential $U_d(x)$,
the OTD approaches a delta function in the $t\to \infty$ limit, $R_0(T_t=T,t)\to \delta\left(T-{{Z_+}\over
{Z}}t\right)$ where
$Z$ is the equilibrium partition function and $Z_+=\int_0^{\infty}e^{-2U_d(x)}dx$ is the restricted
partition function. This result again follows from simple ergodicity arguments. Also, the approach
to this asymptotic form is slow and one finds the
scaling behavior,
$R_0(T_t=T,t)\sim e^{-t\Psi(T/t)}$ in the
scaling limit $T\to \infty$,
$t\to \infty$ keeping $T/t$ fixed where $\Psi(y)$ is the LDF. For the case $F(x)=\mu\, {\rm {sign}}(x)$,
the LDF can be computed exactly and we get $\Psi(y)=2\mu^2(y-1/2)^2$. Recently the LDF has been
computed exactly for other stable potentials also\cite{MB}. In presence of 
disorder $(\sigma>0$),
this asymptotic behavior gets modified as in the case of LTD and we find ${\overline
{R_0(T,t)}}\approx {1\over {t}}f_o(T/t)$ in the scaling limit. The exact calculation of the scaling function
$f_o(T/t)$ is nontrivial but the final answer turns out to be a deceptively simple Beta law,
\begin{equation}
f_o(y)={1\over {B\left(\nu,\,\nu\right)}}{\left[y(1-y)\right]}^{\nu-1}, {\quad\quad 0\le y\le 1}
\label{beta}
\end{equation}
where $\nu=|\mu|/\sigma$ and $B(\nu,\nu)$ is the standard Beta function\cite{GR}. 
If one tunes the parameter $\nu$ by 
either varying $\mu$ or the disorder
strength $\sigma$, this OTD exhibits
an interesting phase transition in the ergodicity of the particle position at $\nu_c=1$. For $\nu<\nu_c$,
the
distribution in Eq. (\ref{beta})
is concave with a minimum at $y=1/2$ and diverges at the two ends $y=0,1$. This means that paths
with small number of zero crossings (such that $T$ is close to either $0$ or $t$)
carry more weights than the paths that cross many times (for which $T$ is close to $t/2$), i.e. the particle
tends to stay on one side of
the origin as in the case of a Brownian motion. Exactly the opposite situation occurs
for $\nu>\nu_c$ where $f_o(y)$ is maximum at its mean value $1/2$ indicating largest
weights for paths that spend equal times on both sides of $x=0$. It is
interesting to notice that similar types of Beta laws also arise in the
study of certain perturbed Brownian motion\cite{CPY}.

$\bullet$ {\it Flat potential} ($\mu=0$)\,: For the pure case ($\sigma=0$),
our method reproduces the
`arcsine' law for the OTD of an ordinary Brownian motion, $R_0(T,t)=1/{\pi \sqrt{T(t-T)}}$. In the presence of
disorder ($\sigma>0$), i.e. for the the Sinai model,
we find that the left part of the OTD for $0\le T\le t/2$ has the large $t$ behavior,
$R_0^{L}(T,t) \approx {1\over {\log t}}R(T)$. The right half of the OTD for $t/2\le T\le t$ is just the
symmetric
reflection of the left part. The $t$ independent function $R(T)$ has a complicated
form but with simple limiting behaviors,
$R(T)\approx \sigma\sqrt{2/{\pi T}}$ as $T\to 0$ and
$R(T)\approx 1/{2T}$ for large $T$, consistent with
the normalization condition $\int_0^{t/2}R_0^L(T,t)dT=1/2$.

In summary we have studied analytically the LTD and OTD for a particle moving in an arbitrary one dimensional
potential
and shown how the quenched disorder changes the asymptotic behaviors
of these distributions. Our exact results are consistent with the general notion that
`disorder broadens distributions' of physical quantities. Recently
several asymptotic exact results for other quantities in the Sinai type models were derived using
a real space renormalization group (RG) treatment\cite{FLM}. Reproducing the exact results presented here
either
via the RG method or by the replica method and extending our results to higher dimensions remain as
challenging open problems.

We thank M. Yor for useful discussions.

\end{multicols}

\end{document}